\documentclass[twocolumn,superscriptaddress]{revtex4-2}

\usepackage[]{graphicx}
\usepackage{dcolumn}
\usepackage{bm}

\usepackage{amsmath,amssymb,amsfonts}
\usepackage{fancyhdr}
\usepackage{lipsum} 
\usepackage{subcaption}
\usepackage{multirow}

 \usepackage[colorlinks]{hyperref}
 \hypersetup{
     colorlinks = false,
   }

\newcommand{\comment}[1]{}

\usepackage{color}
\usepackage{babel}
\newcommand{\bea}{\begin{eqnarray}}
\newcommand{\eea}{\end{eqnarray}}

\newcommand{\be}{\begin{equation}}
\newcommand{\ee}{\end{equation}}

\begin{document}

\author{A. Alonso-Izquierdo}
\affiliation{Department of Applied Mathematics, University of Salamanca, Casas del Parque 2 and \\
Institute of Fundamental Physics and Mathematics,
University of Salamanca, Plaza de la Merced 1, 37008 - Salamanca, Spain}

\author{S. Navarro-Obregón}
\affiliation{Departamento de Física Teórica, Atómica y Óptica, Universidad de Valladolid, 47011 Valladolid, Spain.}

\author{K. Oles}
\affiliation{Institute of Theoretical Physics, Jagiellonian University,
Lojasiewicza 11, Krak\'{o}w, Poland} 

 \author{J. Queiruga}
\affiliation{Department of Applied Mathematics, University of Salamanca, Casas del Parque 2 and \\
Institute of Fundamental Physics and Mathematics,
University of Salamanca, Plaza de la Merced 1, 37008 - Salamanca, Spain}

\author{T. Romanczukiewicz}
\affiliation{Institute of Theoretical Physics, Jagiellonian University,
Lojasiewicza 11, Krak\'{o}w, Poland}

\author{A. Wereszczynski}
\affiliation{Institute of Theoretical Physics, Jagiellonian University,
Lojasiewicza 11, Krak\'{o}w, Poland}

\title{Semi-BPS sphaleron and its dynamics}

\begin{abstract}
We construct a simple field theory in which a sphaleron, i.e., a saddle-point particle-like solution, forms a semi-BPS state with a background defect that is an impurity. This means that there is no static force between the sphaleron and the impurity. Therefore such a sphaleron-impurity system is very much like usual BPS multi-solitons, however, still possessing an unstable direction allowing for its decay.  We study dynamics of the sphaleron in such a system. 
\end{abstract}
\maketitle

\section{Motivation}

Sphalerons are highly nonperturbative static solutions, existing in various nonlinear theories, and to some extent, resembling topological solitons \cite{MS}. They are localized, typically particle-like field configurations obeying vacuum boundary conditions, which, however, do not have to necessarily carry any topological charge. What is important, they are {\it saddle point} solutions i.e., posses an unstable direction along which they decay. Thus their small perturbation spectrum possesses a negative mode. 

Sphalerons have been found in many theories with or {\it without} topological solitons, see e.g., \cite{T, AI, MSam, JMG, KK-1, KS, ST, V, DF}, and for a review see \cite{M}. Therefore, they are more fundamental and generic objects than topological solitons. Of course, the most prominent example is given by the electroweak sphaleron \cite{KM}, see also \cite{K-1, J, KB}. On the contrary to stable topological solitons, which are important for the static or vacuum sectors of a theory (see e.g., instantons and their importance for the structure of the vacuum), sphalerons are crucial also for real time dynamics, e.g., real-time transitions between vacua or scattering processes. In the electroweak theory the sphaleron is responsible for baryon charge violation effects. 

Although sphalerons are known to be very important for time evolution of nonlinear theories, with and without topological solitons, they are much less understood than their stable solitonic counterparts. Specifically, the study of their dynamics or their impact on dynamics of topological solitons is in a rather initial stage. One reason for that  is the complexity of interactions. Indeed, similarly as in the case of solitons, a sphaleron interacts in the three main ways:

First of all, it can be attracted or repelled by sphalerons, solitons or other localized excitations of matter fields. It means that there is a static force acting on a sphaleron due to the presence of defects. This amounts to a kinetic motion. 

Secondly, sphaleron may possess internal degrees of freedom (DoF) i.e., normal, or even quasi-normal modes, which may be excited during scattering processes and which temporary can store some fraction of energy. In the case of topological solitons internal DoF amount to an appearance of fractal structures in the final state formation via the resonant energy transfer mechanism, \cite{CSW, MORW}. In fact, it has recently been shown that sphaleron can provide DoF which trigger resonant transfer phenomenon in kink-antikink collisions \cite{sphaleron}. Hence, they can actively modify solitonic dynamics even though they do not appear in the initial or finite state. 

Thirdly, sphaleron typically meets radiation which also can modify its dynamics.

 Another reason which increases the complexity of the sphaleron dynamics is that it is not an ultimately stable solution and therefore it has a negative, unstable mode. Hence its decay should be also taken into account. Interestingly, this mode is extremely important for the recently discovered {\it sphaleron-oscillon correspondence} \cite{MR} which states that properties of sphaleron and its oscillon (i.e., an oscillon to which this sphaleron decays) are intimately related. This unexpected correspondence partially unifies these two, at the first glance independent, phenomena further underlying the importance of sphalerons in nonlinear dynamics. 

In the case of topological soliton, there exists a very special limit, called {\it Bogomol’nyi-Prasad-Sommerfield (BPS) limit} \cite{B, PS}, which significantly simplifies dynamics and allows for a very nontrivial insight into the static as well as dynamical properties of solitons. In this limit there is no static force between the solitons, which means that the energy of a class of solutions in a topologically fixed sector is degenerated. For example, the constituent solitons, in a multi-soliton sector, can be placed in any spatial point and all such solutions possess exactly the same energy. Here the most famous examples are the abelian Higgs vortices at the critical coupling, the t'Hooft-Polyakov monopoles or the $SU(2)$ Yang-Mills instantons. Importantly, the lowest energy dynamics of solitons in such BPS models finds an elegant formulation in terms of geodesic motion on a pertinent moduli space, which is a space of coordinates parameterizing the energetically equivalent solutions \cite{NM}. No such a BPS-type construction for sphalerons has been studied yet.


The aim of the present work is to propose a framework in which dynamics of a sphaleron does enjoy a sort of BPS limit. We will call such objects {\it semi-BPS sphalerons} to underline their similarities as well as differences in comparison with the typical BPS solitons.  Specifically, we introduce probably the simplest family of theories where a sphaleron remains to be a BPS-like state even in a presence of other localized field configurations, which in our case will be a background, non-dynamical defect i.e., an impurity. This choice is dictated by an obvious reason. Namely, although providing a nontrivial background, which may equivalently be treated as a nontrivial medium, the impurity itself does not introduce new kinetic DoF. This, together with the BPS-like property, may simplify the complexity of the problem allowing for a new insight into the problem of dynamics and interactions of localized solutions. 

However, it should be underlined that, in the case of the sphaleron, being a {\it semi-BPS object} has a slightly different, probably even more nontrivial, meaning. As always, it denotes that:
\begin{enumerate}
 \item There is no static force between such a sphaleron and other localized object (impurity/solitons etc.).
 \item The solution possesses a zero mode (although the impurity explicitly breaks the translational invariance).
 \end{enumerate} 
 On the other hand, a semi-BPS sphaleron does not saturate any topological bound but has a negative mode which can lower the energy of this semi-BPS solution due to a decay along the unstable direction, exactly as required for a genuine sphaleron. Due to that, even in its BPS-like limit the sphaleron reveals much more involved pattern of interactions than its BPS solitonic counterparts. 

We remark that the notion of semi-BPS sphalerons agrees with the concept of semi-BPS solutions, that is solutions which solve a Bogomolny equation on a double-cover of a complex $\phi$-plane \cite{Nick}.

\section{Sphaleron in a scalar field theory}
Let us consider the standard real field theory in (1+1) dimensions
\be
L=\int_{-\infty}^{\infty} \left( \frac{1}{2}\phi_t^2 -\frac{1}{2}\phi_x^2 - U(\phi) \right)dx. \label{Lag-free}
\ee
If the potential has two vacua as for example in the $\phi^4$ theory, 
$U_{\phi^4}=\frac{1}{2} (1-\phi^2)^2$, we have topological solitons, usually called kink (or antikink) interpolating between the vacua,  $\phi_+>\phi_-$. 
An important feature of a kink (or antikink) is that it saturates the pertinent topological energy bound in a given topological sector. Indeed, consider the static energy integral
\bea
E&=&\int_{-\infty}^{\infty} \left( \frac{1}{2}\phi_x^2 +  U(\phi) \right)dx     \\
& =& \int_{-\infty}^{\infty} \left(\frac{1}{\sqrt{2}}\phi_x \mp \sqrt{ U} \right)^2dx \mp  \int_{-\infty}^{\infty}  \sqrt{2U} \phi_x dx \nonumber  \\
&\geq&  \left| \int_{\phi(-\infty)}^{\phi(\infty)}  \sqrt{2U} d\phi \right| = \int_{\phi_-}^{\phi_+}  \sqrt{2U} d\phi \, |Q|, \nonumber
\eea
where the field at infinities must have one of vacuum values. $Q$ is the topological charge, $\pm 1$ for the kink and antikink respectively. This inequality is saturated for solutions of the so-called Bogomolny equations 
\be
\frac{1}{\sqrt{2}}\phi_x = \pm \sqrt{ U}. \label{BOG}
\ee
The kink obeys the equation with plus sign, while the antikink with minus sign. 
\begin{figure}
\includegraphics[width=0.9\columnwidth]{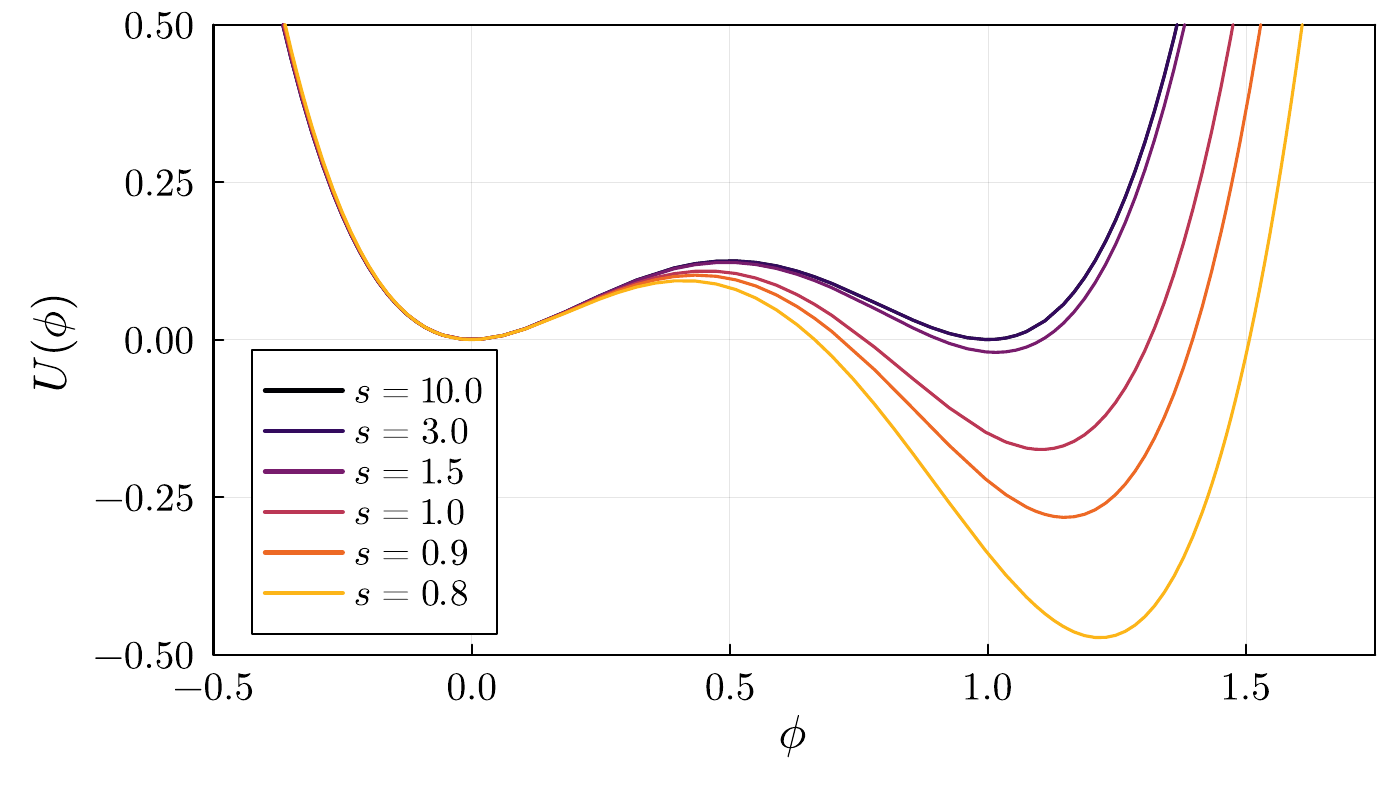}
 \caption{The deformed $\phi^4$ potential $U_s(\phi)$ for different $s$.} \label{potential-plot}
 \end{figure}
\begin{figure}
\includegraphics[width=0.9\columnwidth]{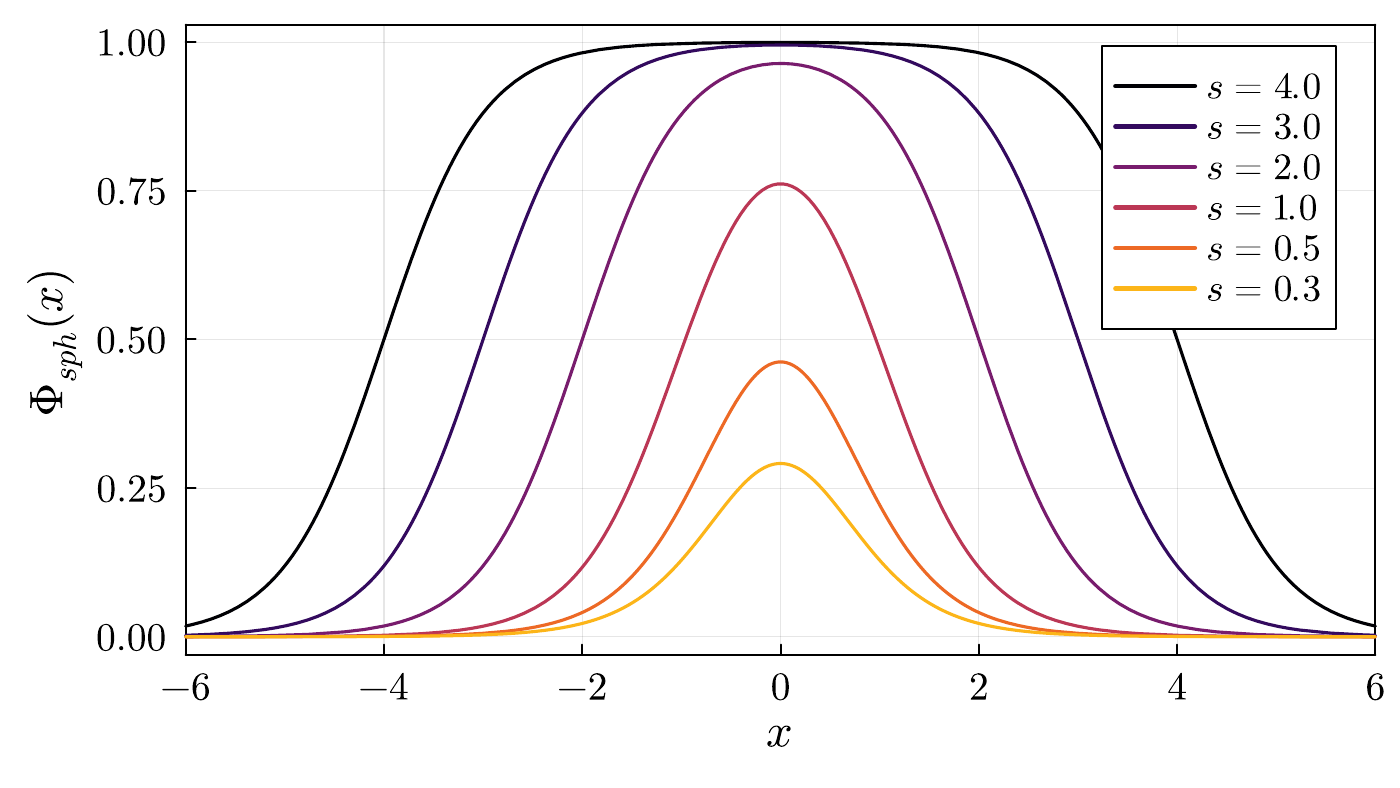}
 \caption{Sphalerons for different values of s in the model $U_s(\phi)$.} \label{sphaleron-plot}
 \end{figure}
 
To find a genuine, spacially localized sphaleron we need a potential where one of the global minima is transformed into a local minimum. Hence, in such models, true vacuum  $\phi_v$ is supplemented by a false vacuum $\phi_f$. Without losing generality we assume that $\phi_v> \phi_f$. In addition, we chose that at the false vacuum the potential takes the zero value, $U(\phi_f)=0$. Since, $U(\phi_v)< U(\phi_f)$, there exists a field value $\phi_0$ at which the potential also vanishes, $U(\phi_0)$. We call this point a {\it return point}. 

Here we make a standard assumption that at the local minimum, $\phi_f$, the potential behaves quadratically while the additional zero $\phi_0$ is achieved linearly i.e., 
\be
U(\phi)= \frac{m_f}{2}(\phi_f-\phi)^2 + O(\phi_f-\phi)^3 \;\; \mbox{at} \;\; \phi \to \phi_f
\ee
and 
\be
U(\phi)= \beta (\phi_0-\phi) + O(\phi_0-\phi)^2 \;\; \mbox{at} \;\; \phi \to \phi_0
\ee
where $m_f$ and $\beta$ are positive constants. A consequence of this is that the false vacuum is approached exponentially at $x \to \pm \infty$ but the zero $\phi_0$ can be approached at a finite point. The extension to a higher order false vacuum is straightforward. 

As an example, which we would like to exploit in the paper, can serve a modified $\phi^4$ potential (e.g., \cite{S, Bazeia})
\be
U_s(\phi)=2\phi^2 \left(\phi-\tanh (s) \right) \left(\phi-\frac{1}{\tanh (s)} \right) \label{pot-sph}
\ee
where $s >0$  is a free parameter. For $s \to \infty$ we recover the $\phi^4$ theory with two true vacua  at $\phi_v=0$ and $\phi_v=1$. Otherwise, there is only one true vacuum at 
\be
\phi_v=\frac{3}{4} \coth (2s) +\frac{1}{4} \sqrt{9 \coth^2(2s)-8}
\ee
and one false vacuum (local minimum) at $\phi_f=0$ at which $U_s(\phi_f)=0$, see Fig. \ref{potential-plot}. In the limit $s\to 0$ we effectively recover the $\phi^3$ model where the true vacuum is sent to infinity \cite{MR}. This potential has a quadratic zero at the false vacuum and a linear zero at the return point $\phi_0=\tanh (s)$. Another linear zero at $\phi=\coth(s)$ is not important for our considerations.  

This model admits a sphaleron which interpolates between the false vacuum and passes through the additional zero of the potential \cite{Bazeia}. Its exact form reads
\be
\Phi_{sph} (x)= \frac{1}{2} \left( \tanh(x+s) - \tanh(x-s)\right). \label{sph}
\ee
It may be viewed as a superposition of the $\phi^4$ kink and antikink located at $\mp s$ respectively, see Fig. \ref{sphaleron-plot}. For sufficiently large $s$ we easily recognize both $\phi^4$ kink and antikink in the sphaleron profile. For $s \to 0$ the solution is approximately $1/\cosh^2(x)$ which, up to some rescalings, is the $\phi^3$ sphaleron \cite{MR, CC, Bazeia}. 

 In fact all potentials with a false vacuum amount to the appearance of such a sphaleron. 
Interestingly, it may be treated as a solution which  {\it piece-wisely} obeys the Bogomolny equation. Namely, 
\be
\frac{1}{\sqrt{2}}\phi_x= \left\{
\begin{array}{cc}
   +\sqrt{U}  & x\leq 0 \\
   & \\
    -\sqrt{U} & x\geq 0.
\end{array} 
\right.
\ee
At the gluing point (which here is set at the origin) the field takes $\phi_0$ value. Therefore these two branches are connected forming a $\mathcal{C}^1$ solution. In fact, one can easily show that all derivatives are continuous at this point, which results in a smooth $\mathcal{C}^\infty$ sphaleron. This can be approached in a very elegant way in terms of integration theory for kinks and sphalerons \cite{Nick}. 

The fact that sphalerons in scalar field theories fulfill piece-wisely the pertinent Bogomolny equation is a guideline which will be exploited in the next part of the paper.

Of course, to prove that this is a sphaleron one has to compute the linear perturbation spectrum and identify a negative, unstable mode. This is indeed the case for our choice of the potential. The easiest way to verify this is to compute the zero mode and show that it possesses one node. Therefore there must be a lower (negative) energy mode. 

Needless to say that, since the theory is translational invariant, the center of the sphaleron can be moved from the origin to any spatially point $x_0$. One may view this sphaleron as a trivial BPS-like object. There is a (translation) zero mode but its action on the solution is trivial. For example, the spectral structure is $x_0$ independent. 
\section{Non-chiral BPS-impurity models}

Typically, in single scalar field theory both kink or sphaleron, if placed in the presence of other (antik)kink or sphaleron, are subject to an attraction or repulsion. This also means that such a multi-particle state usually does not have any zero mode. As a consequence, it does not obey any Bogomolny-type equation. Hence, it is not a BPS solution.

To find multi-kink BPS solutions one needs to add, in a very special manner, at least one new scalar field \cite{F, Izq}, which obviously, makes investigations more complicated both from analytical as well as numerical point of view. However, it was recently shown that it is possible to couple the scalar field $\phi$ to a background, non-dynamical field $\sigma(x)$ in such a way that BPS property is preserved \cite{BPS-imp-1, BPS-imp-2}. It means that one obtains the simplest BPS-type theory, still with only one dynamical field. Indeed, here a soliton does not statically interact with the other localized object, i.e., the impurity and therefore they can be located at any distance from each other without change in the energy. 
\begin{figure*}
\includegraphics[width=0.9\columnwidth]{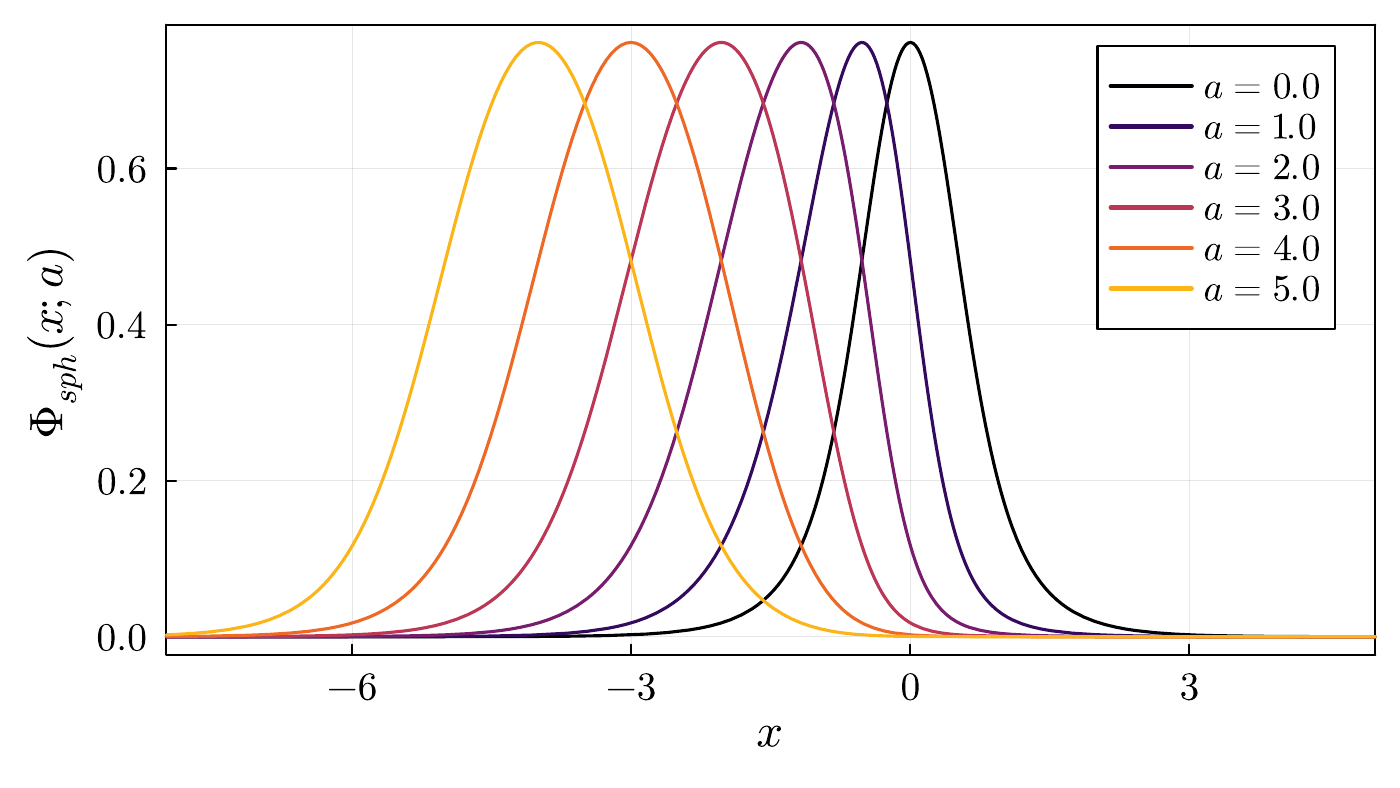}
\includegraphics[width=0.9\columnwidth]{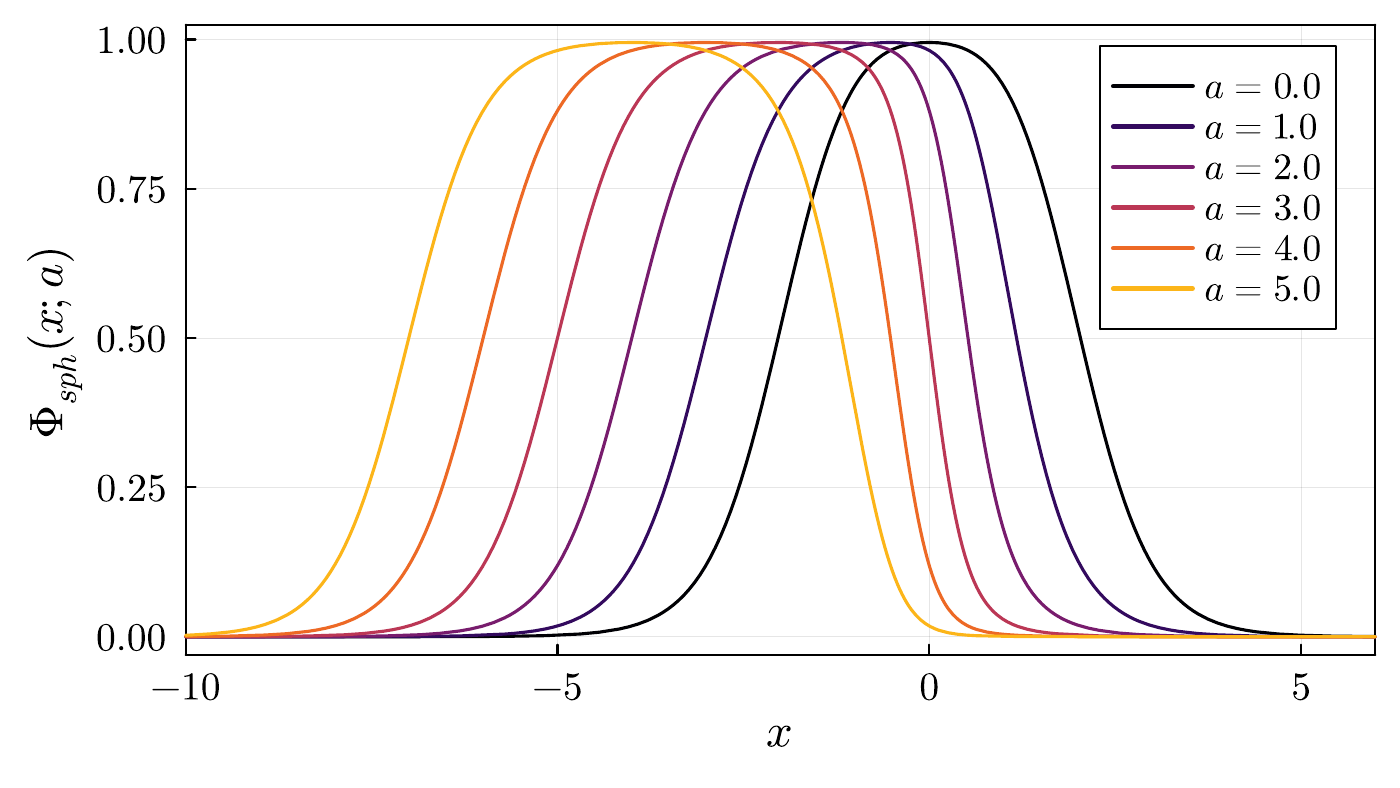}
\includegraphics[width=0.9\columnwidth]{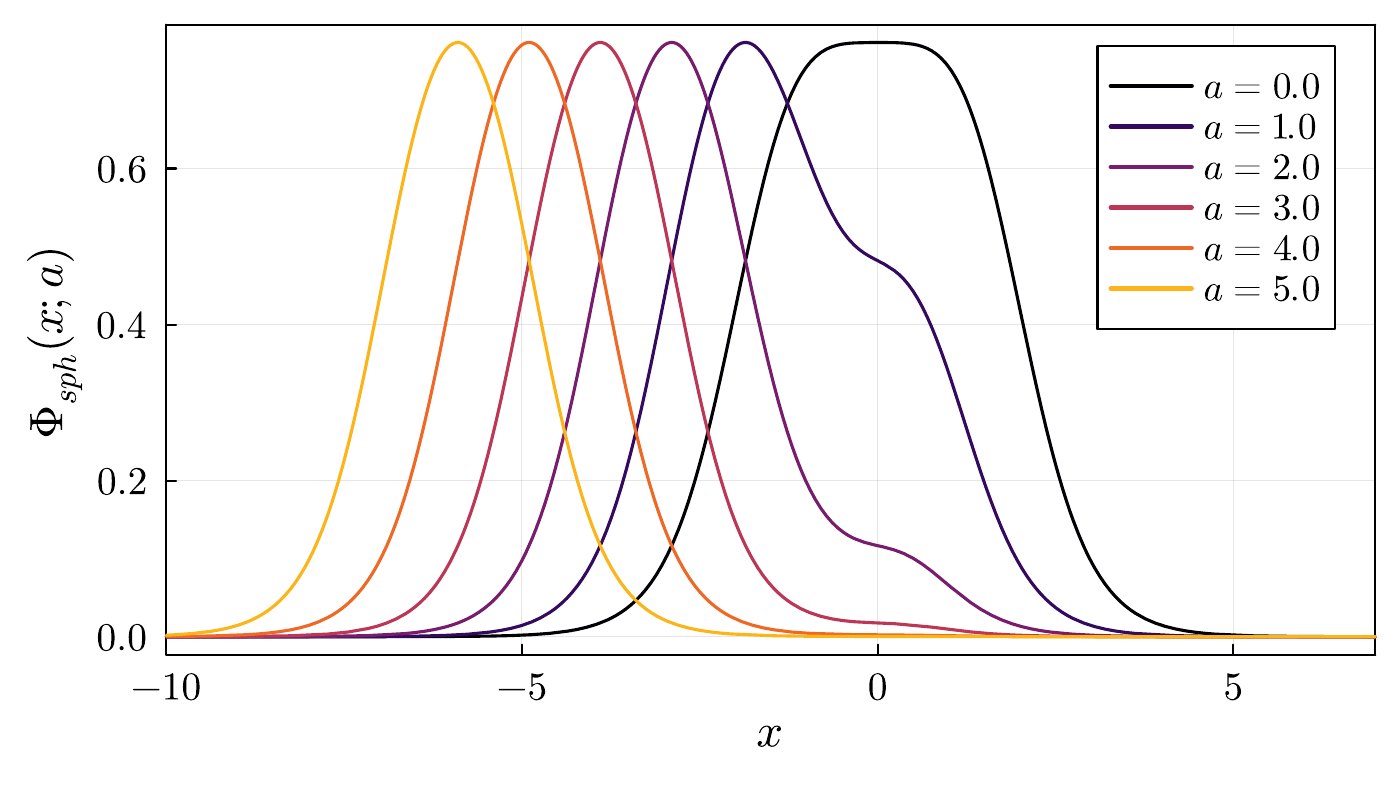}
\includegraphics[width=0.9\columnwidth]{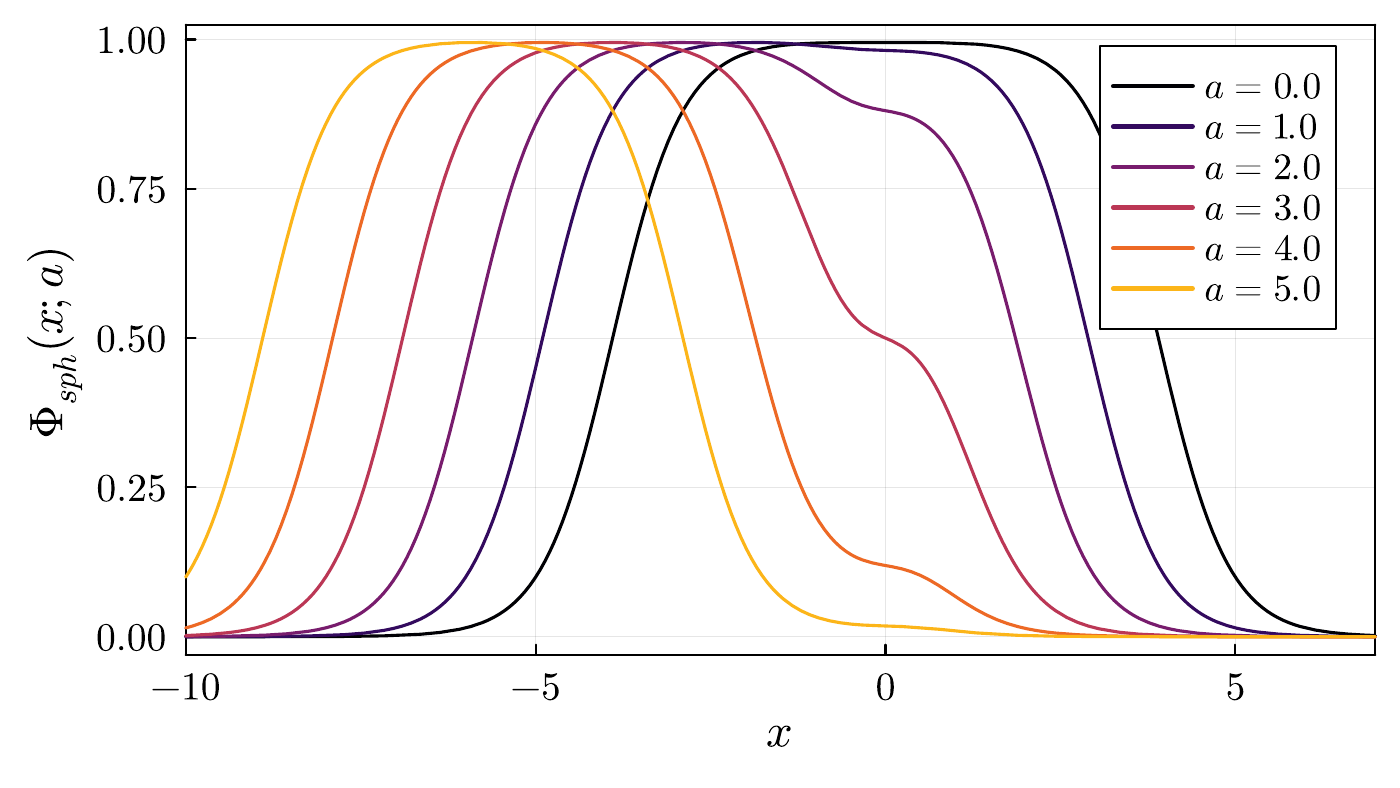}
 \caption{The BPS sphaleron-impurity solution for different values of the modulus $a$. Upper panels: $\alpha=1$, lower panels $\alpha=-0.9$. Left panels: $s=1$, right panels: $s=3$.} 
 \label{sph-imp-plot}
 \end{figure*}

The usual BPS-impurity modification changes {\it one} of the Bogomolny equations (\ref{BOG}) into the following equation \cite{BPS-imp-1, BPS-imp-3}
\be
\frac{1}{\sqrt{2}} \phi_x= \sigma(x) W(\phi), 
\ee
where $W$ is a well-behaved function of $\phi$ and $W^2=U$. 
 
Such a modification of the Bogomolny equation can be achieved if the original field theory is deformed as  \cite{BPS-imp-1}
\be
L=\int_{-\infty}^{\infty} \left( \frac{1}{2}\phi_t^2 -\frac{1}{2}\left( \phi_x - \sigma(x) W(\phi) \right)^2 \right)dx.
\ee
Again, the existence of a first order Bogomolny equation implies that there is a family of energetically degenerate kink-impurity solutions which are parameterized by a continuous parameter $a$, called {\it modulus} and which may be related to a distance between the kink and the impurity. In comparison to the no impurity case, (\ref{BOG}), the BPS sector is nontrivial. For example, the linear mode structure of the soliton-impurity BPS solution depends on the distance between the soliton and impurity. This renders the BPS-impurity model an ideal, simplified laboratory for studying the role of the internal modes in kink dynamics in the limit of vanishing static force. In fact, it allowed for the discovery of the spectral wall phenomenon \cite{AORW} which only very recently has been observed in various solitonic models without any impurity \cite{two-field-SW, SW-phi6}. The BPS-impurity framework also allows for the computation of one-loop corrections to kink-antikink processes, see \cite{TW}.
 
Importantly, there is {\it only one} Bogomolny equation in this deformed model. Hence, the resulting soliton can be treated as a chiral kink. Based on the previous analysis, where we underlined the importance of the existence of two Bogomolny equations, we may conclude that, while there are BPS kink-impurity or even BPS kink-antikink-impurity solutions, no BPS sphaleron-impurity states  are possible in this set-up. 

To construct a model which possesses a BPS sphaleron-impurity solution we have to modify BPS-impurity Lagrangian in such a way that two Bogomolny equations show up. It means that we need a non-chiral BPS-impurity theory.

For that reason let us consider the following background field deformed Lagrangian
\be
L=\int_{-\infty}^{\infty} \left( \frac{1}{2}\phi_t^2 -\frac{1}{2\sigma}\phi_x^2 - \sigma U(\phi) \right)dx, \label{BPS-imp-L}
\ee
where again $\sigma$ is a given impurity. But now the background field modifies the gradient term, too. To find Bogomolny equations we again analyze the static energy integral
\bea
E&=&\int_{-\infty}^{\infty} \left( 
\frac{1}{2\sigma}\phi_x^2 + \sigma U(\phi) \right)dx 
  \\
& =& \int_{-\infty}^{\infty} \left(\frac{1}{\sqrt{2\sigma}}\phi_x \mp \sqrt{\sigma U} \right)^2dx \mp  \int_{-\infty}^{\infty}  \sqrt{2U} \phi_x dx \nonumber\\
&\geq&  \left| \int_{\phi(-\infty)}^{\phi(\infty)}  \sqrt{2U} d\phi \right| = \int_{\phi_-}^{\phi_+}  \sqrt{2U} d\phi \, |Q|. \nonumber
\eea
The inequality is saturated if and only if
\be
\frac{1}{\sqrt{2\sigma}}\phi_x = \pm \sqrt{\sigma U}. \label{BOG-imp} 
\ee
This proves that such a simple impurity model has two Bogomolny equations and therefore does not treat kink or antikink differently. Note, that the impurity cannot be a function with negative value but otherwise it is not restricted by any fine-tuned condition. Similarly as before a sphaleron-impurity BPS solution is a solution which pice-wisely solves the impurity modified Bogomolny equations (\ref{BOG-imp}). This requires a potential with false vacuum, e.g., (\ref{pot-sph}). 

In fact, the impurity dependent Bogomolny equation can be rewritten in the impurity free form after introducing a new base space coordinate $y$
\be
\frac{dy}{dx}=\sigma(x) \; \Rightarrow \; y=\int \sigma dx + a
\ee
where $a$ is an integration constant.  Then, we arrive at the usual (no impurity) Bogomolny equations
\be
\phi_y = \pm \sqrt{2 U}. 
\ee
Solving these equations and replacing the $y$ coordinate in terms of the original coordinate $x$ leads to a family of BPS sphaleron-impurity solutions. Note, that this solution is parameterized by a modulus $a$ reflecting the BPS property of the solution. 

Let us now turn to our potential (\ref{pot-sph}) and for concreteness assume the following form of the background field
\be
\sigma(x)=1+\frac{\alpha}{\cosh^2(x)},
\ee
which is an exponentially-like localized impurity centered at the origin. $\alpha > -1$ is a free parameter measuring its strength. For $\alpha=0$ the impurity vanishes and we recover the usual theory (\ref{Lag-free}).  For our choice of the impurity the sphaleron-impurity solution reads 
\bea
\Phi_{sph}(x; a)&=& \frac{1}{2} \left( \tanh(x+\alpha\tanh(x)+a+s) \right. \nonumber \\
&-& \left. \tanh(x+\alpha\tanh(x)+a-s)\right), \label{sph-imp}
\eea
where the modulus $a$ can be related to the distance between the sphaleron (position of the maximum of the field) and the impurity (origin). Indeed, for $|a|\gg 1$, we get the undeformed sphaleron far away from the origin (impurity). However, as $a$ tends to 0, the sphaleron approaches the impurity and deforms. For positive $\alpha$ the original sphaleron gets squeezed as approaching the impurity, while for $\alpha \in (-1,0)$ it is widened. As we approach the limiting value $\alpha \to -1$ derivative of the field at the origin tends to 0. This behaviour is presented in Fig. \ref{sph-imp-plot}. 

We remark that the sphaleron-impurity solution enjoys a reflection symmetry
\be
\Phi_{sph}(-x;-a)=\Phi_{sph}(x;a).
\ee

All these solutions possess exactly the same $a$-independent energy 
\be
E[\Phi_{sph}(x;a)] = \frac{9 \sinh(2s) + \sinh(6s) - 24 s \cosh(2s)}{6 \sinh^2(2s)} 
\ee
and form a BPS-type family of sphalerons in the sense explained before. The sphaleron can be placed at any distance from the localized impurity without changing its energy. Thus, there is no static force between the sphaleron and impurity. 

\vspace*{0.2cm}

In the subsequent part of the work we will analyze various aspects of dynamics of such a semi-BPS sphaleron-impurity system. 

\section{The mode structure}
 \begin{figure}
\includegraphics[width=1.0\columnwidth]{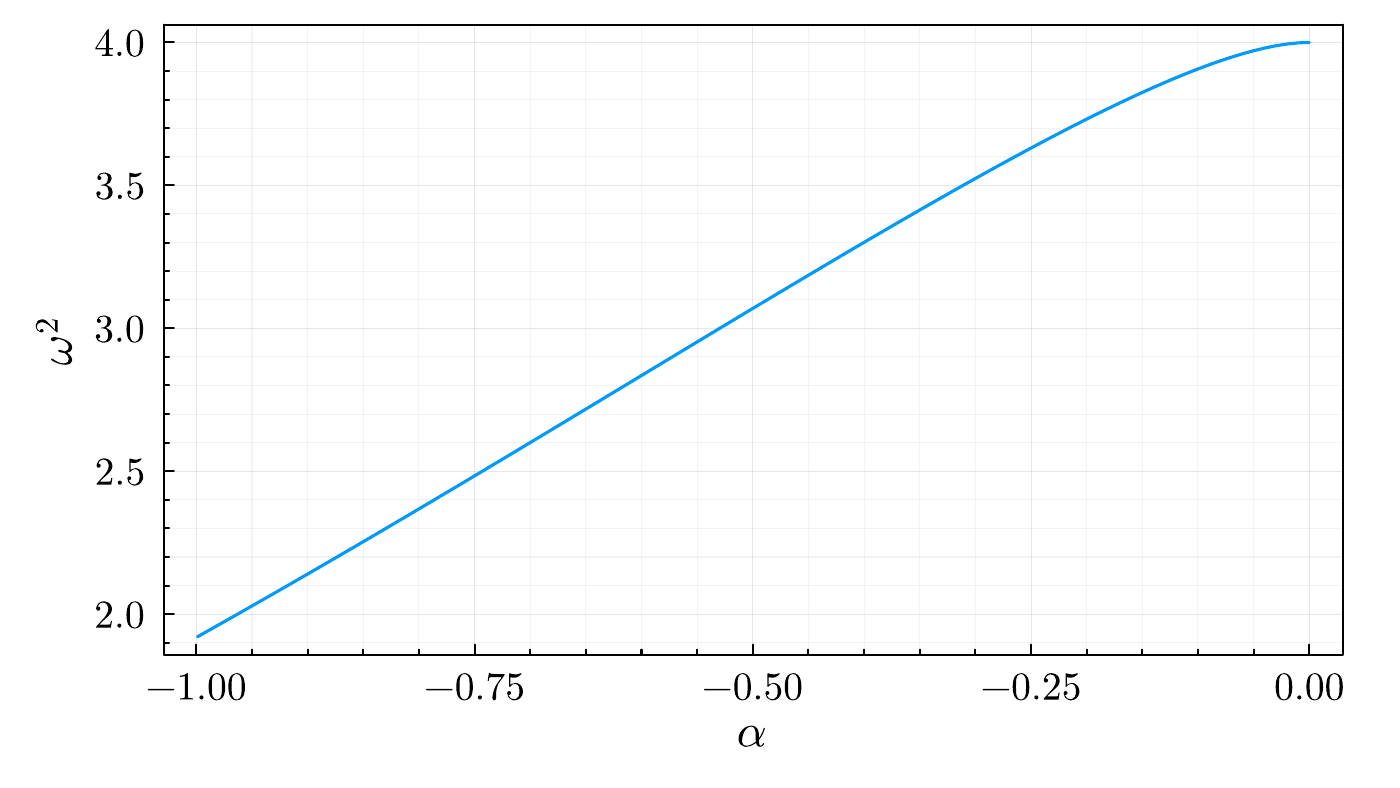}
 \caption{The frequency of the bound mode of the impurity. }
 \label{imp_flow}
 \end{figure}

Although the energy is degenerate while we change the modulus $a$ it is not the case for the linear mode structure. This is related to the fact that the impurity is not just a uniform vacuum and therefore nontrivially modifies the shape of the sphaleron. To see that we consider the static BPS solution with an addition of a small perturbation $\eta(x,t)$
\be
\phi(x,t)=\Phi_{sph}(x;a)+\eta(x,t)
\ee
and insert it into the full time-dependent field equation 
\be
\phi_{tt}-\frac{1}{\sigma}\phi_{xx}+\frac{\sigma_x}{\sigma^2} \phi_x + \sigma U_\phi=0.
\ee
At linear order we get
\bea
&& \omega^2(a)\eta (x;a)  =  \\
&& \left[- \frac{d}{dx} \left( \frac{1}{\sigma} \frac{d}{dx}  \right) + \sigma U''(\Phi_{sph}(x;a))\right] \eta(x; a), \nonumber
\eea
where we assume that $\eta(x,t;a)=e^{i\omega(a) t} \eta(x;a)$. Here, $\omega(a)$ is frequency of the pertinent mode $\eta(x;a)$. Dependence on the modulus $a$ is explicitly marked. 

Asymptotically, for $a \to \pm \infty$, the semi-BPS sphaleron $\Phi_{sph}(x;a)$ is just an infinitely separated system of the sphaleron in the original (no impurity) model $\Phi_{sph}(x)$ and the impurity. Thus, the mode structure is just a superposition of the modes of the free sphaleron and modes which are added due to the background field itself. The former ones are localized on the sphaleron while the later ones are localized on the impurity (here at the origin). In our example, the free sphaleron, at least for not too small $s$, is just a pair of well separated kink and antikink and therefore it has two positive discrete modes $\eta_{1,2}(x)$ originated by the shape modes of each of the constituent solitons. Thus, as $s$ becomes larger their frequencies tend to $\omega^2_{1,2}=3$. There is also a zero mode $\eta_0(x) \sim \partial_x \Phi_{sph}(x)$. Finally, there  is the negative mode, $\eta_{-1}(x)$ whose frequency tends to 0 for $s \to \infty$. In other words, the sphaleron becomes quasi-stable for large $s$.

As we already noted the impurity may add some additional modes. This is found by linear perturbation around $\phi=0$ solution. In our example there is only one such a mode localized at the origin and it arises for $\alpha <0$, see Fig. \ref{imp_flow}. This plot is independent of the value of the model parameter $s$. It follows from the fact that $U''(\phi=0)=4$. This mode will actively participate in dynamics of the sphaleron. 

\begin{figure}
\includegraphics[width=1.0\columnwidth]{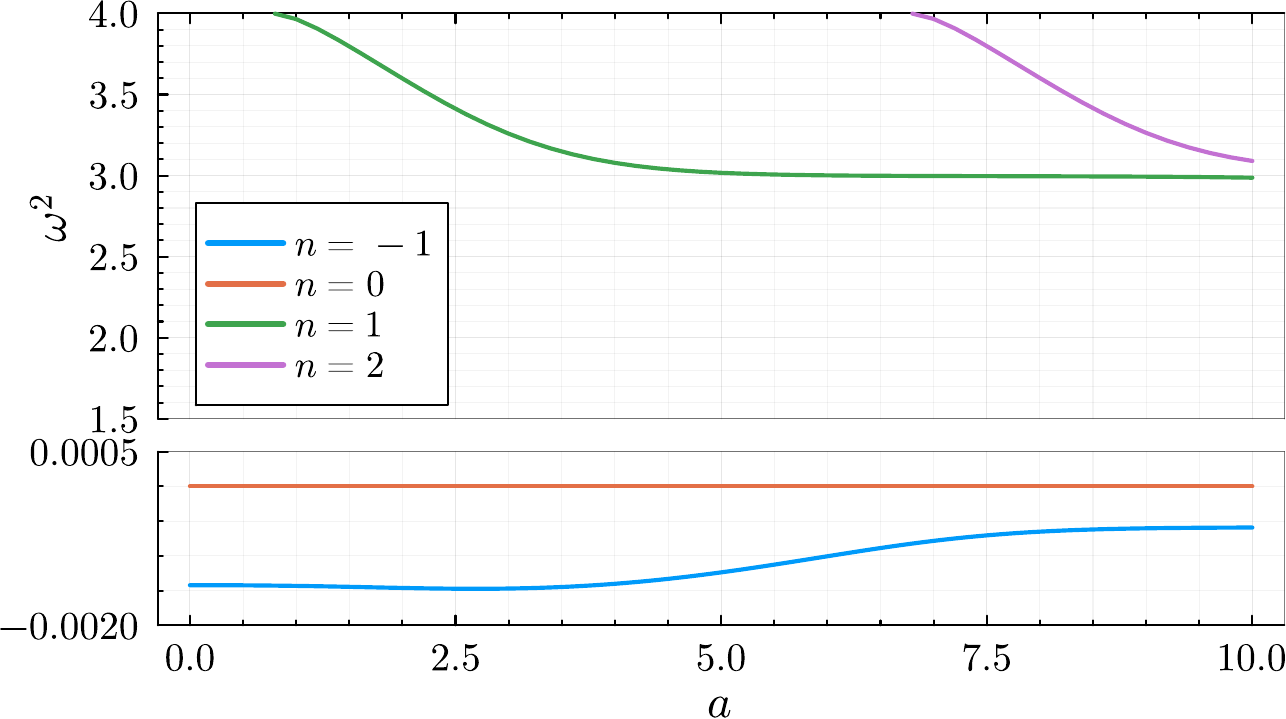}
\includegraphics[width=1.0\columnwidth]{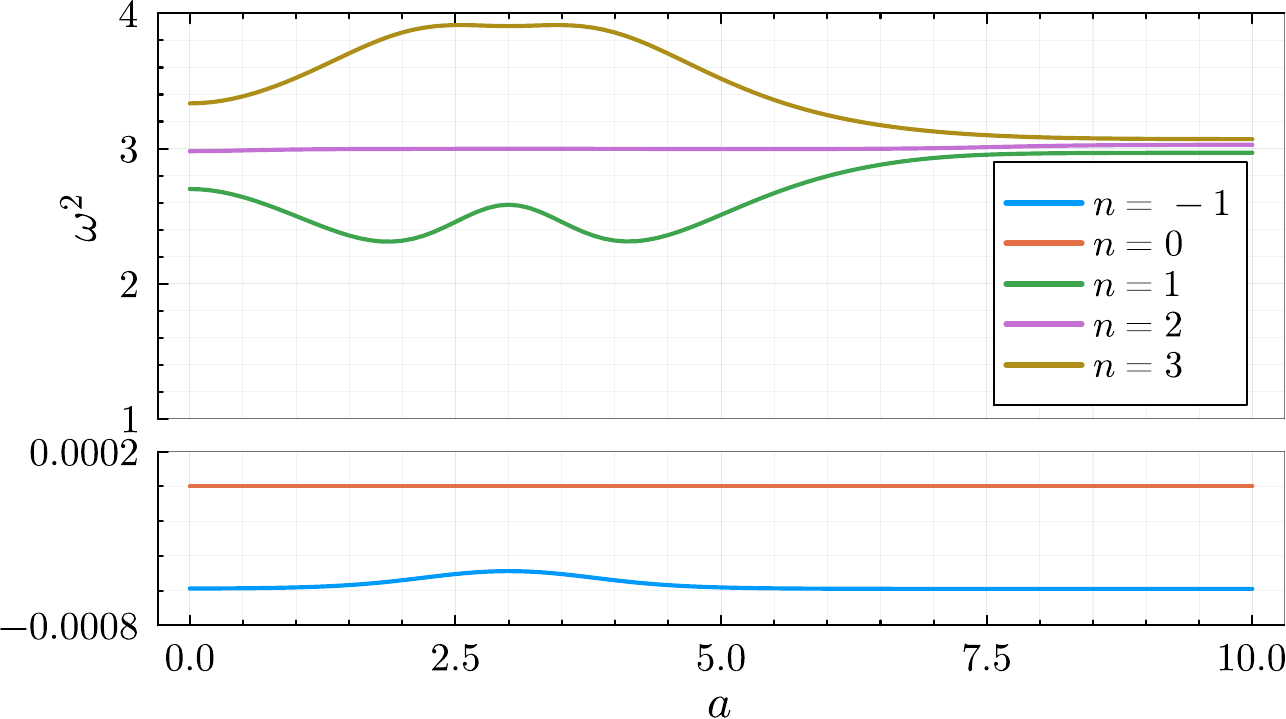}
 \caption{The dependence of the mode structure of the sphaleron-impurity solution on the modulus $a$. Upper panel: $s=3$, $\alpha=3$. Lower  panel: $s=3$, $\alpha=-0.5$. }
 \label{spectral_flow}
 \end{figure}

If the sphaleron approaches the impurity it gets deformed and the spectral structure changes in a more and more significant manner. 
Of course, due to the BPS nature of the solution the zero mode still exists even though the translational symmetry is lost
\be
\eta_0(x;a)= N(s,\alpha) \; \partial_a \Phi_{sph}(x;a),
\ee
where the normalization constant $N(s,\alpha)$ is fixed for a given potential and background field., i.e., for given $s$ and $\alpha$. Importantly, the zero mode has one node. This implies that the solution is a genuine sphaleron with one negative energy, unstable mode. In general, however, the spectral structure may be deformed in a rather drastic way, see Fig. \ref{spectral_flow}. 

First of all, the background field can have impact on the unstable mode. Namely, as $a\to 0$, the frequency of the unstable mode frequency rises/ lowers for negative/positive strength $\alpha$, respectively. Thus, approaching the origin the sphaleron becomes more or less stable depending on the impurity. 

Secondly, also the spectrum of the massive bound modes changes. For sufficiently large positive impurity the bound modes of the free sphaleron can cross the mass threshold at a certain $a=a_{sw}$. This fact is known to trigger the spectral wall phenomenon which has a significant impact on kink dynamics \cite{AORW}. In section \ref{sec-SW} we will study this phenomenon closer.

Finally, we remark that there is an interesting possibility related to the change of the frequency of the negative mode. By an appropriate choice of the impurity a genuine, fast decaying sphaleron can be transformed into a quasi-stable sphaleron with the negative mode very close to 0. This happens at a certain distance from the impurity due to the sphaleron-impurity interaction.
\section{The geodesic dynamics } 
 
In the well known BPS soliton limit, BPS solutions saturate the pertinent topological bound. This means that in a given topological sector there are no solutions with lower energy. A natural consequence of that is that the simplest, lowest energy dynamics occurs via transitions between energetically equivalent BPS solutions, which reflects an excitation of the zero mode. This found an elegant formulation as a geodesic motion on the corresponding moduli space of the BPS states \cite{NM}. 
 
In the case of the BPS sphalerons we do have a family of energetically degenerate solutions $\mathcal{M}[a]=\{ \Phi_{sph}(x;a)\}$ and therefore the concept of geodesic motion may still make sense. However, the sphaleron is not the global energy minimizer in its, here trivial, topological sector. Hence, in the space of field configurations, there is an unstable direction along the negative mode $\eta_{-1}(x;a)$. Therefore looking from this perspective the BPS sphaleron solutions span a sort of unstable moduli space. 
\begin{figure}
\includegraphics[width=0.9\columnwidth]{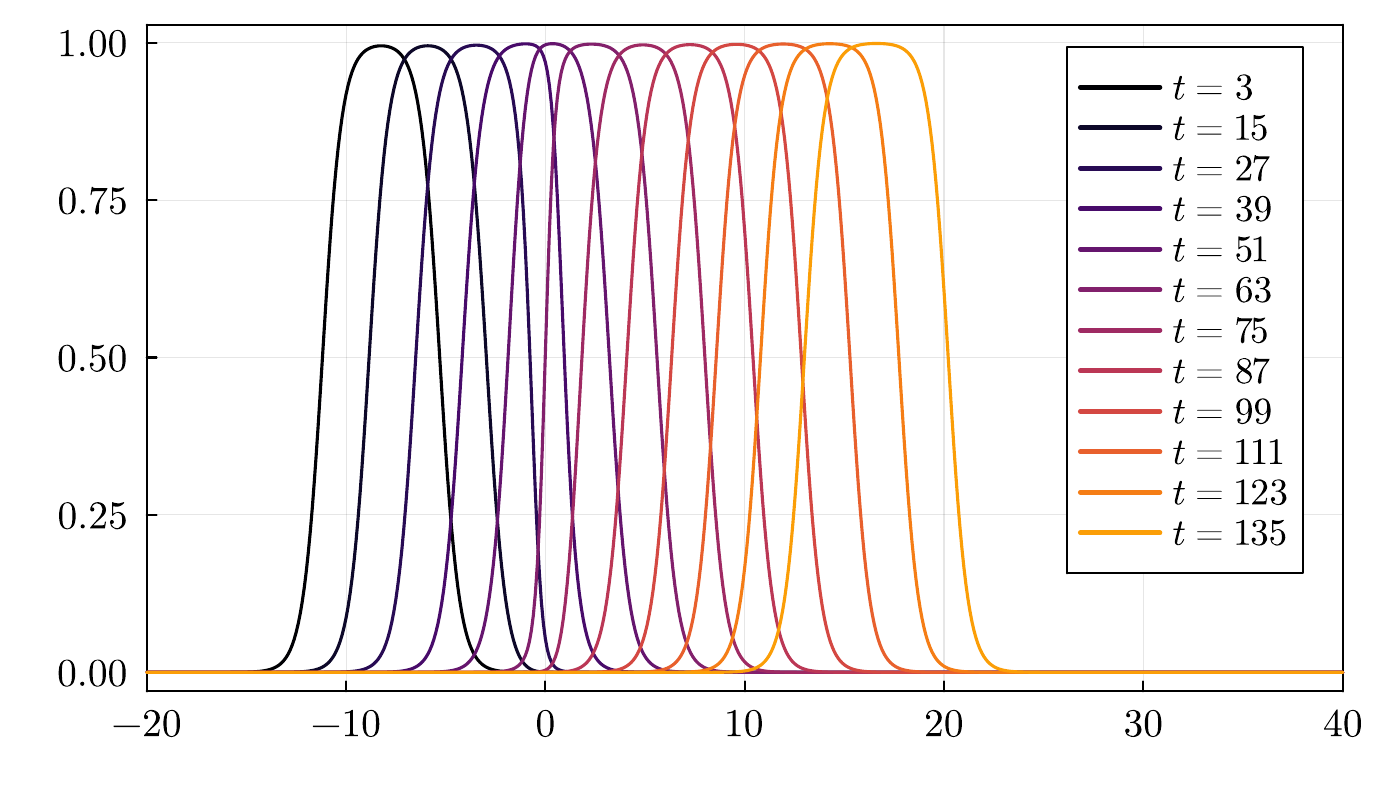}
\includegraphics[width=0.9\columnwidth]{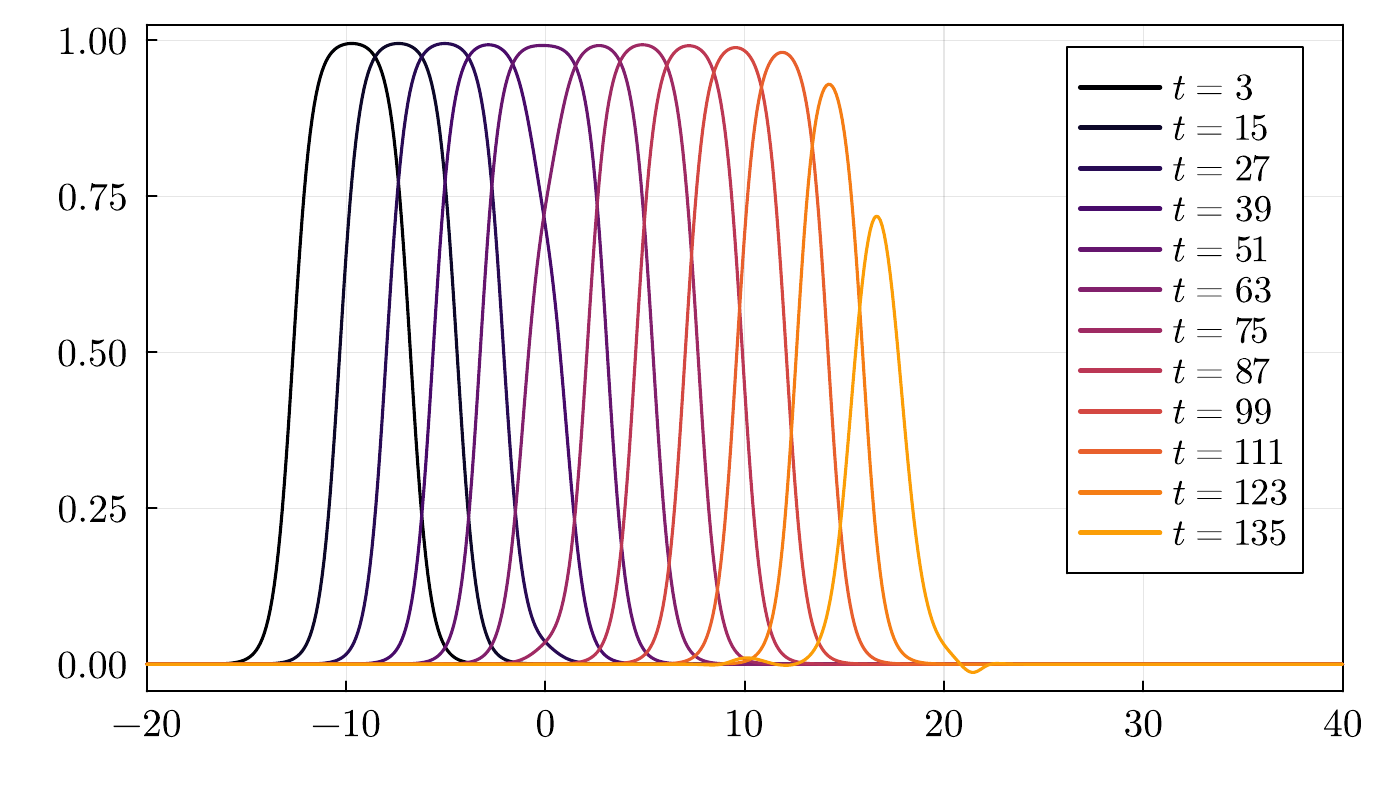}
 \caption{Collision of the sphaleron ($s=3$) for repulsive ($\alpha=1$) and attractive ($\alpha=-0.5$) BPS impurity, with $v_{in}=0.195$.} \label{sph-geo-dyn-plot}
 \end{figure}
 
As always, the metric on such a moduli space is defined as \cite{MS, M} 
\be
g(a)=\int_{-\infty}^\infty \left( \frac{d\Phi_{sph}(x;a)}{da} \right)^2 dx
\ee
This can be obtained by inserting the BPS sphaleron-impurity  solutions into the original field-theoretical Lagrangian and promoting the modulus $a$ to a time dependent collective coordinate $a(t)$. Such an insertion gives rise to a one-dimensional collective coordinate model (CCM)
\be
L[a]= \frac{1}{2} g(a) \dot{a}^2, 
\ee
whose solution is just a geodesic flow on the moduli space $\mathcal{M}[a]$. We omitted a dynamically unimportant constant potential term. Under particular circumstances the unstable mode may remain unexcited also when the sphaleron passes through the impurity. This frequently occurs when the sphaleron is a quasi-stable solution which happens for sufficiently large $s$. Then, one can expect that, for a long time, the evolution follows the geodesic flow on the unstable moduli space $\mathcal{M}[a]$. This is precisely what we saw in our numerical analysis. 

\begin{figure}
\includegraphics[width=0.9\columnwidth]{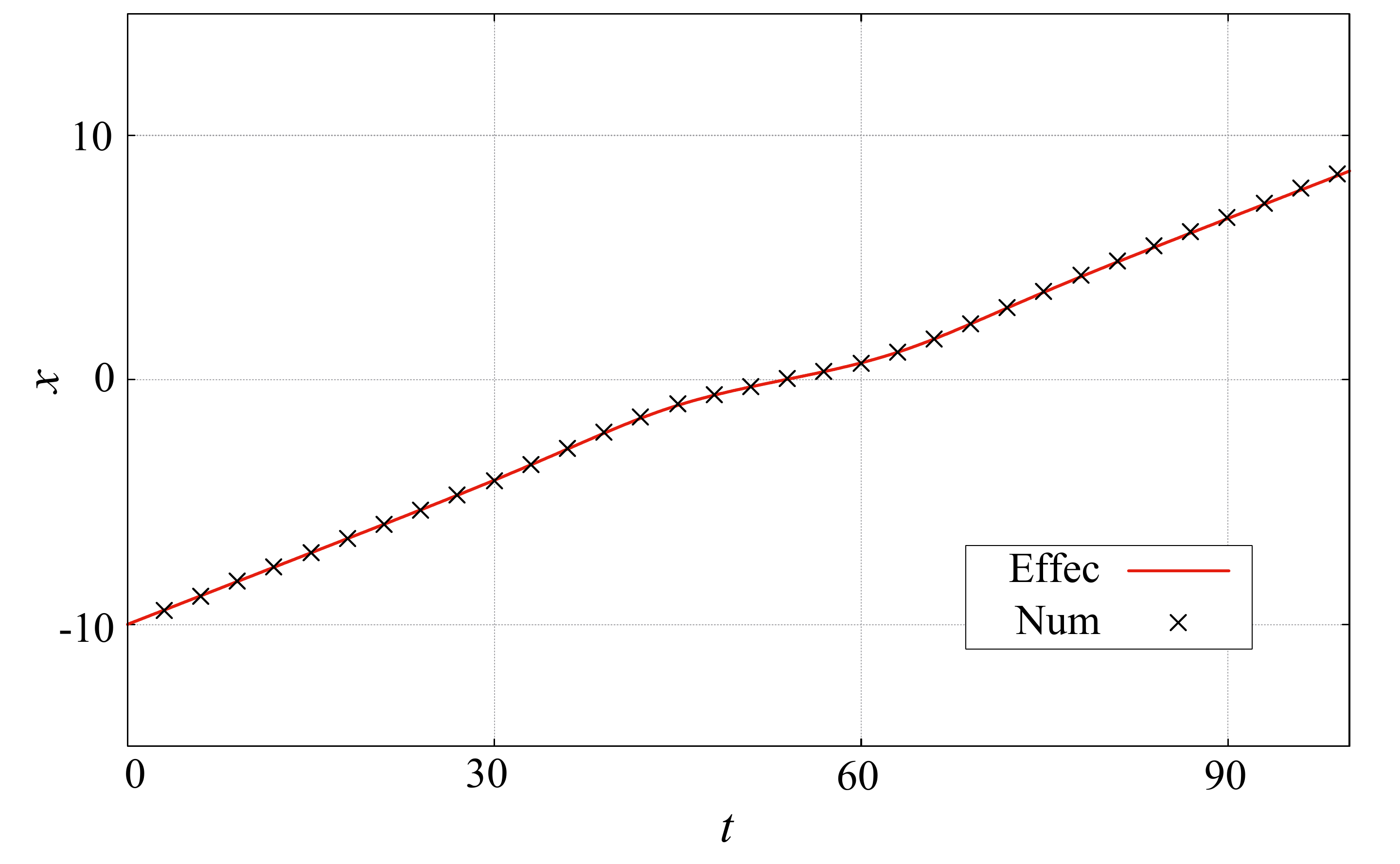} 
\caption{Comparison between full numerics and the moduli space flow dynamics for a non-decaying boosted sphaleron. The parameter of the model are $s = 3, \alpha = 2$. The initial velocity is $v = 0.15$.}
\label{fig:geodesic_flow}
\end{figure}

Specifically, in Fig. \ref{sph-geo-dyn-plot} we show profiles of the field while the sphaleron passes through the impurity. In the upper panel, for $s=3$ and $\alpha=2$, the sphaleron is distorted by the impurity exactly as dictated by the moduli space analysis. The initial velocity of the sphaleron is $v_{in}=0.15$. The profiles at fixed time $t$ are consistent with the BPS solutions at some value of the modulus $a$. Furthermore, actual full field theory dynamics  is also very well reproduced by the CCM geodesic flow, see Fig.  \ref{fig:geodesic_flow}.

However, the sphaleron by definition possesses an unstable mode and it may be excited, due to higher order nonlinear effects, while the sphaleron goes through the impurity. 

In Fig. \ref{sph-geo-dyn-plot}, lower panel, we present an example of such a process where after passing the impurity the sphaleron destabilizes and starts to decay. Whenever the unstable mode gets excited then the geodesic flow on the $\mathcal{M}[a]$ space breaks down and the sphaleron falls apart. 

To conclude this section, the geodesic motion based on the CCM model may still quite well describe the simplest dynamics of a semi-BPS sphaleron but, as one might expect, it is a less powerful tool if compared with BPS solitons. 

\section{Spectral wall } 
\label{sec-SW}

Now we will go beyond the simplest geodesic dynamics and investigate how excitation of massive bound modes changes the interaction between the semi-BPS sphaleron and impurity. In particular, we are interested in the case where a bound mode of the static sphaleron-impurity solution crosses the mass threshold at some $a$. It is known that such a property, if existing in a BPS kink theories, gives rise to the spectral wall phenomenon. 

Speaking precisely, a spectral wall phenomenon is an obstacle in soliton dynamics due to a transition of a mode into the continuum spectrum. This happens for a certain value of a moduli space coordinate $a=a_{sw}$ which can be translated to a distance $x_{sw}$ between the soliton and its collision partner (another soliton or impurity). The effect strongly depends on the amount of the mode exited on the soliton. If its amplitude $A$ is smaller than a critical value $A<A_{crit}$ then the soliton passes $x_{sw}$ with a distortion growing with increasing of the amplitude. We say that it passes the spectral wall. If $A>A_{crit}$ then the soliton gets reflected by the spectral wall where the reflection point occurs sooner for bigger $A$. Finally, if $A=A_{crit}$, the soliton forms a very long living quasi-stationary state at a fix position $x=x_{sw}$.  

Interestingly, the spectral wall phenomenon has a very selective nature. Namely, each mode which crosses the mass threshold has its own spectral wall. Furthermore, position of the spectral wall can be at very large distance from the interaction partner. Thus, it can trigger a long range interaction. 

To study this issue in the context of sphaleron we take, as an initial configuration, a boosted semi-BPS sphaleron with one of its massive modes $\eta(x;a)$ excited. We also assume that at $t=0$ the sphaleron is at a large distance $x_0$ from the impurity. Hence, at $t=0$ 
\bea
\phi(x,t)&=&\Phi_{sph}(\gamma(x-vt+x_0))\\
&+&A \eta(\gamma(x-vt+x_0)) \cos (\omega \gamma (t-vx)) \nonumber
\eea
which provides Cauchy data for the numerical analysis. Of course, the included mode must hit the mass threshold for some $a$ (i.e., for some distance between the sphaleron and origin). This is the case for a sufficiently big strength $\alpha$ of the impurity. Furthermore, for simplicity we will search for spectral walls in the quasi-stable sphaleron limit, that is for a large $s$. In particular, in our example, we choose $s=8$ and $\alpha=3$. Due to that, the free sphaleron looks like a molecule of kink and antikink of $\phi^4$ theory with a large separation between the constituent solitons. Hence, its massive modes are very well approximated by a symmetric and antisymmetric superposition of the shape modes of the $\phi^4$ kinks
\begin{figure}
\includegraphics[width=1.0\columnwidth]{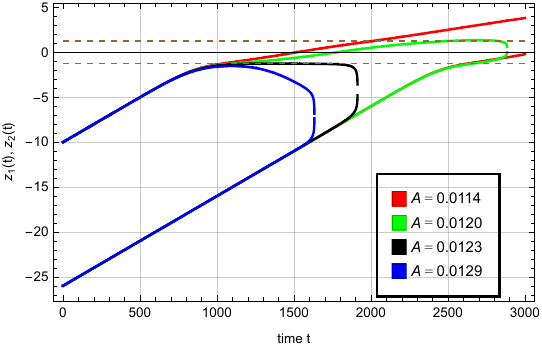} 
\caption{The collision of the sphaleron with a spectral wall. The lines show positions $z_1,z_2$ of the composite kink and antikink in the sphaleron. The colors denote cases with different amplitude of the excited mode. Dashed lines are the positions of the spectral wall.}
\label{fig:SW}
\end{figure} 
\be
\eta_{\pm} (x) = N_{\pm} \left( \eta_{sh} (x+s) \pm  \eta_{sh} (x-s)  \right),
\ee
where
\be
\eta_{sh}(x) \approx \frac{\sinh(x)}{\cosh^2(x)},
\ee
is the usual shape mode of the $\phi^4$ kink and $N_{\pm}$ are normalization constants. The initial velocity of the sphaleron is $v_{in}=0.01$. In Fig. \ref{fig:SW} we present the dynamics of the sphaleron with the $\eta_+$ mode excited. From the perspective of the composite anti-kink the corresponding spectral wall is located at $x_{sw}=1.25$. Indeed, at this distance from the origin the mode hits the mass threshold. We clearly see in our numerics the appearance of the spectral wall. At the critical value of the amplitude of the mode the composite antikink forms a stationary state at $x=x_{sw}$. 

The same pattern is observed for smaller $s$. However, for $s<4$ addition of the shape mode responsible for the spectral wall should be compensated by an addition of unstable mode to prevent the sphaleron to collapse \cite{MR}. 

Obviosuly, such a spectral wall acts as another destabilizing factor which contributes to the decay of the sphaleron.

\section{Decay of the semi-BPS sphaleron} 

By definition, the sphaleron is an unstable solution, here, with two decay channels. The first is a trivial one and describes a decay into the true vacuum. The sphaleron splits into kink and antikink which continually accelerate in opposite directions. The second possibility is much more interesting. It corresponds to the decay into the false vacuum and creation of an oscillon. However, this oscillon is not a BPS-type object. Furthermore, it is not immersed in the trivial vacuum but in a rather complicated background provided by the impurity. Hence, its patterns of interactions are much more involved than in the usual non-impurity model. 

For simplicity, we analyze the decay of sphaleron when it is located at the origin. Then, the problem maintains the reflection symmetry. However, one should be aware that decay of the semi-BPS sphaleron depends on its position with respect to the impurity.

One interesting scenario occurs if the impurity possesses its own internal mode. Then, the sequence of events is the following. First, the sphaleron decays into its oscillon which oscillates in the background of the impurity. As a consequence, the oscillon loses energy, reduces its amplitude and increases its frequency. After quite a long time the oscillons settle down at a mode of the impurity which finally slowly decays via emission of radiation. So, we may conclude that in this case the oscillon is trapped by the impurity and transformed into a bound mode. 
\begin{figure}
\includegraphics[width=1.0\columnwidth]{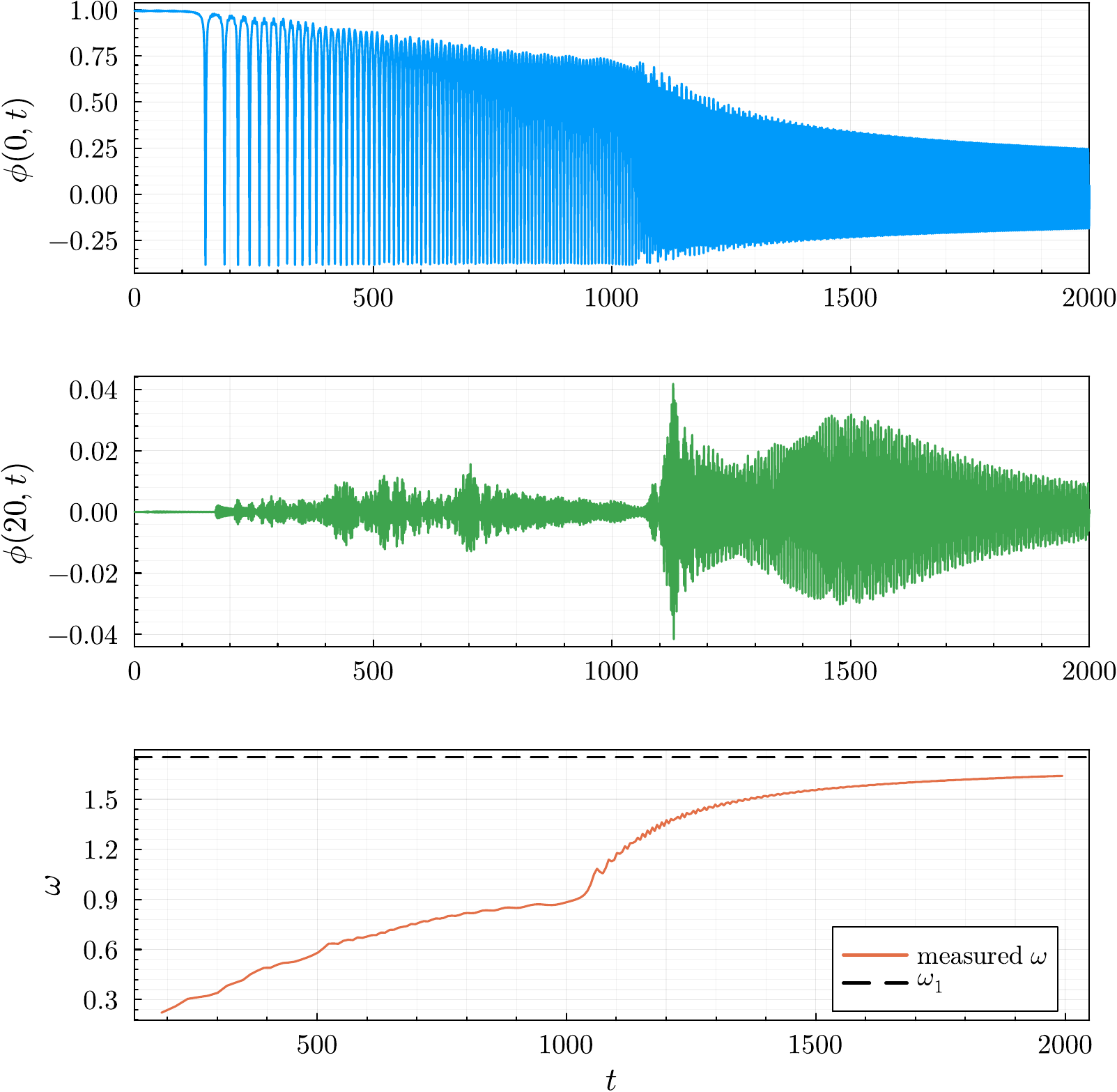}
 \caption{Decay of the sphaleron for $s=3$ and $\alpha=-0.5$. } \label{StoO-1}
 \end{figure}
 
In Fig. \ref{StoO-1} we present an example of such a behavior for $s=3$ and $\alpha=-0.5$. In the first phase, for $t < 1050$, the large oscillon gradually lowers its amplitude, see the upper panel. As time grows, we also see a characteristic double oscillations of the amplitude. Suddenly at $t \approx 1050$ there is a rapid jump both in the amplitude and frequency, see the upper and bottom panel. The reason for that is the radiation burst effect. It arises due to the fact that the frequency of oscillations is half of the mass threshold (m=2) and therefore the double harmonics may freely propagate. The increase of radiation is clearly visible in the central panel, where we present the field measured at $x=20$. After this moment of the evolution the system tends to a configuration being the impurity with a mode excited. This is a very slow process which we see particularly well if a dumping is added to the field equation. 
 \begin{figure}
\includegraphics[width=1.05\columnwidth]{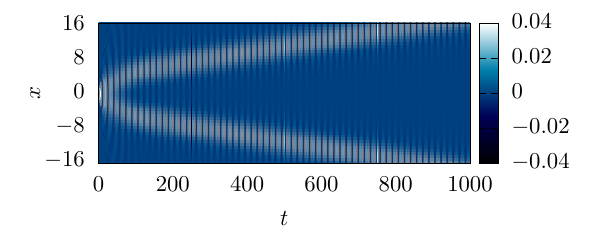}
\includegraphics[width=1.05\columnwidth]{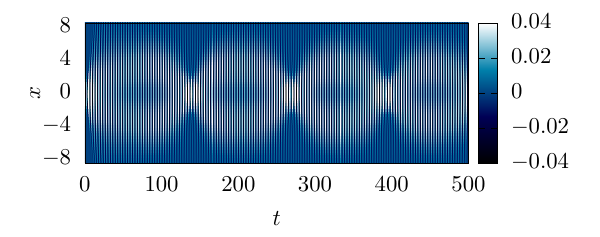}
\includegraphics[width=1.05\columnwidth]{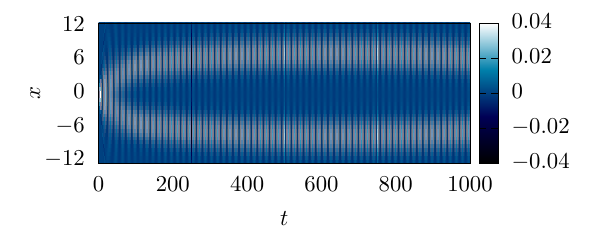}
 \caption{Decay of the sphaleron and interaction of the emerging oscillon with the impurity: $\phi(x,t)$ for $s=0.2$. Upper: $\alpha=0.0595$ - ejection of the oscillons; central: $\alpha=0.0575$ - bouncing of oscillons; lower: $\alpha=0.05925$ - formation of a quasi-stationary state. } \label{StoO-2}
 \end{figure}
 
Several comments are in order. Firstly, initially the sphaleron reappears at the turning points of the oscillon oscillations. We can also see that the sphaleron carries some excitations of its internal mode, see upper panel and small wiggles in Fig. \ref{StoO-1} when $\phi(0)$ is close to $\phi=1$.

 Secondly, one can notice that radiation in the large oscillon phase is much lower than in the phase where the internal mode starts to participate in the decay of the oscillon.
 
 Thirdly, the existence of a mode of the impurity effectively looks as appearance of an attractive force between the oscillon and impurity. In our example, this is indeed the case for a negative $\alpha$. 
 
\hspace*{0.2cm}

A drastically different scenario happens if the impurity does not host any bound mode. The sphaleron again decays into its oscillon which evolves in the background of the impurity. After a short time this oscillon is destabilized by the impurity and splits into two oscillons which are ejected in opposite directions, see Fig. \ref{StoO-2} upper panel, where $s=0.2$ and $\alpha=0.0595$. Hence, we observe a repulsive interaction between the oscillon and impurity. In our example it occurs for positive $\alpha$. 

However, a detailed analysis shows that the situation is much more involved. The interaction between the ejected oscillons and the impurity possesses also an attractive channel. An example of that evolution is presented Fig. \ref{StoO-2} central panel. We see that both ejected oscillons turn back and bounce around the origin. Here $s=0.2$ and $\alpha= 0.0575$. 

Interestingly, we discovered a limiting situation where these oscillons form a long living quasi-stationary state. Indeed, for a very long time their positions stay fixed at a certain distance from the impurity. This is clearly visible in Fig. \ref{StoO-2} lower panel, where $s=0.2$ and $\alpha=0.05925$.
\section{Semi-BPS sphaleron-impurity collisions} 

As we already pointed out, once the sphaleron decays, the system completely loses its semi-BPS property and the resulting oscillon reveals a rather nontrivial pattern of interaction with the impurity. This is precisely observed if the semi-BPS sphaleron is scattered with the impurity. The main outcome of such processes is the creation of the oscillon in the first collisions and its subsequent interaction with the impurity. 

\begin{figure*}
\includegraphics[width=2.1\columnwidth]{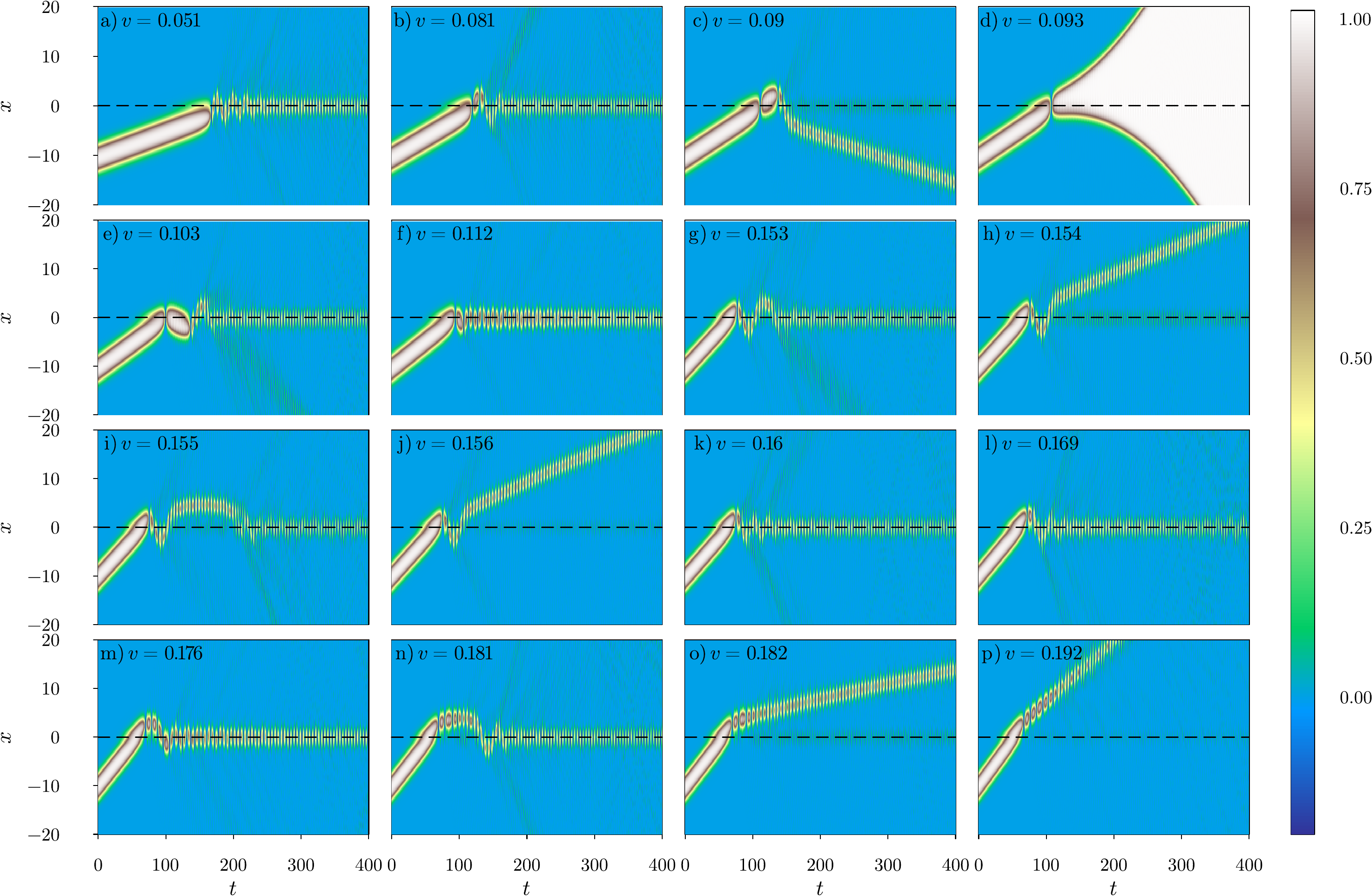}
 \caption{Examples of scenarios for the sphaleron impurity collisions for $s=2.5$ and $\alpha=-0.5$ for increasing initial velocity: from 0.051 to 0.192} \label{S-I-scan-1}
 \end{figure*}

In Fig. \ref{S-I-scan-1} we present an example of collisions of a quasi-stable sphaleron, $s=2.5$, with an attractive impurity, $\alpha=-0.5$. For smaller velocities, $v_{in}< v_1\approx 0.051$, the sphaleron is always destabilized by the impurity and decays into an oscillon. Then the oscillon is trapped by this attractive impurity exactly as described in the previous section i.e., it excites a bound mode of the impurity. On the other hand, if the initial velocity is larger than $v_{2}\approx 0.182$, then the sphaleron passes through the impurity. This again destabilizes the sphaleron which eventually, at a certain distance from the impurity, decays into a boosted oscillon
 
However, between these two cases there is a variety of different processes. We define the following qualitatively distinct classes of behaviour:
\begin{enumerate}
    \item {\it Bounces of the oscillon.} 
    \\
    Sphaleron decays into oscillon which is then attracted by the impurity. As a consequence, the oscillon performs oscillations around the impurity (e.g., panels a, b, g, k and l in Fig. \ref{S-I-scan-1}).  
    \item {\it Oscillon trapping or ejection.} 
    \\
    Such oscillations release energy and after a few bounces we end up with trapping of the oscillon eventually leading to an excited impurity (e.g., panels a, b, f, g, k, l and m)  or ejecting the oscillon both in the forward (panels h and j) and in the backward direction (panel c).
    \item {\it Oscillon stationary state.}
    \\
    The oscillon can also form a stationary state where it stays at a constant distance from the impurity (panel i). This case resembles very much the stationary state found in decay of a semi-BPS sphaleron in the presence of repulsive impurity. 
    \item {Reappearance of sphaleron.} 
    \\
    Sphaleron collapses but then it is created again (panels c and e). After the creation it can decay into the oscillon or into a pair of kink and antikink which escape from each other (panel d). 
\end{enumerate}

\begin{figure}
\includegraphics[width=1.05\columnwidth]{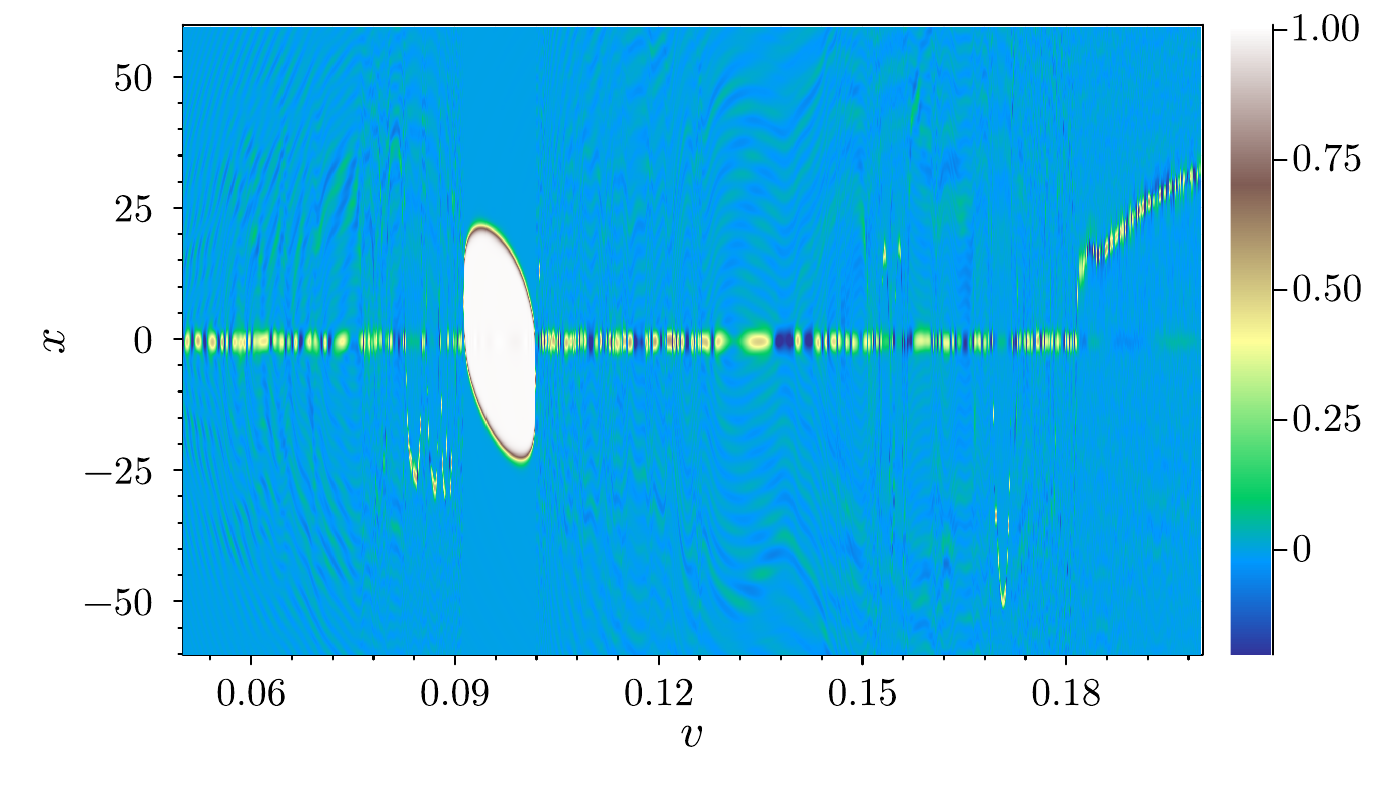}
 \caption{The spatial dependence of the field $\phi(x)$ at large time for different initial velocity $v_{in}$. } \label{SI-final}
 \end{figure}

By varying the initial velocity of the sphaleron we found an evidence that the formation of the final state reveals a structure of a probably chaotic nature. Indeed, different scenarios follow one after the other in a rather random way, see Fig. \ref{SI-final}, where we plot the field at the large time after collision. One can easily see that the trapping of the oscillon by the impurity dominates. Other possibilities like backward and forward ejection of the oscillon as well as the kink-antikink creation, are also clearly visible. 
\section{Summary} 

The main achievement of the present work is establishing the existence of a type of BPS sphaleron-impurity solutions. We call them {\it semi-BPS sphalerons} because of the fact that they comprise two rather different features. 

On the one hand, they resemble the usual BPS systems known from the theory of topological solitons, which means the existence of a family of energetically equivalent static solutions which obey a lower order differential equation. This results in the appearance of a moduli space with a coordinate being a continuous parameter (modulus) parameterising the solutions and in existence of a zero mode which is not related with the trivial translation invariance of the theory. Specifically, in our setup, the sphaleron can be put at any distance from the impurity and the total energy of the system remains unchanged. This means that there is no static force between sphaleron and impurity.  

On the other hand, there is a fundamental difference in comparison with BPS solitons. Namely, the semi-BPS sphaleron does not saturate any topological energy bound and therefore there is an unstable direction along which such a static solution can decay. In other words, there is a negative mode in the spectrum of small perturbations. This is, of course, an expected property of a genuine sphaleron. 

As pointed out by Manton \cite{Nick} the semi-BPS property means that the solutions obey the corresponding Bogomolny equation on a double-cover of the complex plane of $\phi$. 

After proving the existence of semi-BPS sphaleron-impurity solutions we investigated their dynamical properties. The reason is obvious. As in BPS multi-soliton cases, the absence of a static force between localized objects which participate in scatterings (here the sphaleron and impurity) provides a simplified environment and allows for a deeper insight into dynamical properties of sphalerons. Indeed, we found that the semi-BPS sphaleron impurity system reveals many properties known form the usual BPS multi-soliton systems. 

For example, in the simplest dynamics, i.e., where only the kinetic DoF is excited, the semi-BPS sphaleron follows the {\it geodesic motion} on the pertinent moduli space, which means that during such a time evolution the field passes through the available BPS states. 

Next, if one excites a positive bound mode, the sphaleron may be subject to the {\it spectral wall} phenomenon. As always, the necessary condition is that such a mode touches the mass threshold at certain value of the modulus, which translates into a certain distance between the sphaleron and impurity. Such a spectral wall seems to always destabilize sphalerons leading to their collapse, mainly to an oscillon. This was clearly observed in the case where the sphaleron can be treated as an unstable bound state of kink and antikink. It is similar to the effect played by spectral walls in some supersymmetric BPS-impurity models where they destroy bound state of the bosonic and fermionic DoF \cite{Ferm}. 

We remark that there is a range of the model parameter,  where the sphaleron looks like an almost stable kink-antikink molecule. Here it occurs for sufficiently large $s$. Hence, it is natural to expect that collisions of such {\it quasi-stable} sphalerons may find an interpretation as a four-kink scattering.

This standard BPS-soliton like dynamics is strongly modified once the unstable mode is excited. Then the field probes a direction in the configuration space along which the sphaleron can decay forming e.g., an {\it oscillon}. Of course, due to the saddle point nature of the sphaleron the unstable mode can be excited by an arbitrarily small perturbation and therefore, after sufficient amount of time, it will always be the main factor of the evolution. Importantly, the oscillon is not a BPS object and therefore interacts with the impurity in a rather sophisticated manner. In this sense the semi-BPS sphaleron is a less powerful concept than the usual BPS limit. 

Nonetheless, our framework still allows for the study of the sphaleron decay in a simplified, BPS-like situation i.e., in the limit of the absence of the sphaleron-impurity force. Simultaneously, the produced oscillon is immersed in a very nontrivial environment which amounts to appearance of an intriguing {\it oscillon-impurity stationary state}. Its origin, as well as understanding its properties and importance requires further studies. 

The current work can be in a natural way extended to models where the background field is replaced by a dynamical field. The resulting field theory will contain at least two coupled scalar fields. Some of such models indeed host a family of energetically equivalent unstable solutions. See for instance the Montonen-Sarker-Trullinger-Bishop (MSTB) model \cite{Mo, STB}, where a one-parameter family of sphalerons (in this case, non-topological solutions connecting and returning to the same point in a two-dimensional internal space) exists \cite{Izq}. They could decay into one of the vacua, possibly giving rise to radiation emission and the presence of oscillons, but they could also decay into two less energetic kinks. It would be interesting to check how the dynamical properties of sphalerons and oscillon found in our work look in this model. Sphalerons can also be found in the Bazeia-Nascimento-Ribeiro-Toledo (BNRT) model \cite{BNRT} for certain values of the coupling constant. In this case, they are unstable topological kinks that can decay into other less energetic kinks \cite{IGG}. From our perspective, the study of these processes is also relevant. Further examples of theories with sphalerons can be found along the line described in \cite{HB}. 

Interestingly, there is another type of BPS sphalerons in higher dimensions. These again are energetically equivalent unstable static solutions with moduli space parametrized by a continuous parameters. Hence they possess at least one zero mode. On the contrary to the semi-BPS sphalerons considered here they do not solve the original Bogomolny equations. The nonlinear (2+1) dimensional $\sigma$-models with $\mathbb{CP}^n$ and $F_ 2$ target space provides the most studied cases, \cite{Z-1, Z-2, AS}. It would be desired to understood a possible relation between these sphalerons and semi-BPS ones. One should also remark that due to the scale symmetry these sphalerons cannot support any bound modes and therefore their dynamical properties can be quite different. 

A more general question is whether any sphaleron can be understood as an unstable multi-soliton state. Here it is indeed the case. For large $s$ we clearly appreciate a kink-antikink inner structure. In fact, for {\it all} $s$ considered here, the sphaleron can be written as a superposition of a kink and antikink. Interestingly, such a picture reproduce the unstable and the zero mode of the sphaleron (and possible also the shape modes). Namely, the zero mode is related to a shift of the positions of the constituent kink and antikink in the same direction, $z_{1,2} \to z_{1,2}+b$ while the unstable mode corresponds to a similar shift but in opposite direction, $z_{1} \to z_{1} + b$, $z_{2} \to z_{2}-b$. It is clear that the second transformation changes the distance between the constituents.

A related problem is how properties of sphalerons are affected by external conditions. This includes, for example, the appearance of external magnetic and electric fields or other localized objects (solitons or sphalerons). As we saw, for some impurities, the frequency of the negative mode comes closer to 0, which may transform a rather unstable sphaleron to a quasi-stable version. We note that impact of a nontrivial background (e.g., external magnetic field) on properties of sphalerons has a great physical importance \cite{HR}. 


\section*{Acknowledgements}

K. O., T. R., and A. W. were supported by the Polish National Science Centre (Grant No. NCN 2019/35/B/ST2/00059). A.A.I., S.N. and J.Q. were supported by the Spanish Ministerio de Ciencia e Innovación (MCIN) with funding from the  European Union NextGenerationEU (PRTRC17.I1) and the Consejería de Educación, Junta de Castilla y León, through QCAYLE project, as well as MCIN Project No. PID2020-113406GB-I00 MTM. 
S.N. thanks JCyL for the pre-doctoral contract financed through ESF+.

K.O. and A. W. thank Nick Manton for inspiring discussions and for sharing an early version of \cite{Nick}. We also thank Azadeh Mohammadi for comments.

\end{document}